\documentclass{aa}
\usepackage{times}
\usepackage{latexsym}
\usepackage{epsfig}
\def\kms{km~s$^{-1}$}
\def\0{\hspace*{0.5em}}

\begin{document}

\title{The origin of the runaway high-mass X-ray binary
HD153919/4U1700-37\thanks{Based on data obtained with ESA's
astrometric satellite {\it Hipparcos}.}}

\author{A. Ankay\inst{1,2} \and L. Kaper\inst{1} \and J.H.J. de
Bruijne\inst{3} \and J. Dewi\inst{1} \and R. Hoogerwerf\inst{3} \and
G.J. Savonije\inst{1}}

\offprints{L. Kaper;  e-mail: lexk@astro.uva.nl}

\institute{Astronomical Institute ``Anton Pannekoek'' and Center for
           High-Energy Astrophysics, 
           University of Amsterdam, Kruislaan 403, 1098 SJ Amsterdam, 
           The Netherlands
           \and Middle East Technical University, Physics Department,
           06531, Ankara, Turkey
           \and Sterrewacht Leiden, Leiden University, Postbus 9513,
           2300 RA Leiden, The Netherlands}

\date{Received; Accepted}

\authorrunning{Ankay et al.}
\titlerunning{The runaway system HD153919/4U1700-37}

\abstract{Based on its {\it Hipparcos} proper motion, we propose that
the high-mass X-ray binary HD153919/4U1700-37 originates in the OB
association Sco~OB1. At a distance of 1.9~kpc the space velocity of
4U1700-37 with respect to Sco~OB1 is 75~\kms. This runaway velocity
indicates that the progenitor of the compact X-ray source lost about
7~M$_{\odot}$ during the (assumed symmetric) supernova explosion. The
system's kinematical age is about $2 \pm 0.5$ million years which
marks the date of the supernova explosion forming the compact object.
The present age of Sco~OB1 is $\la 8$ Myr; its suggested core,
NGC~6231, seems to be somewhat younger ($\sim 5$ Myr). If
HD153919/4U1700-37 was born as a member of Sco~OB1, this implies that
the initially most massive star in the system terminated its evolution
within $\la 6$ million years, corresponding to an initial mass $\ga
30$~M$_{\odot}$. With these parameters the evolution of the binary
system can be constrained.
\keywords{Stars: early type -- Stars: mass loss
-- Stars: neutron -- Stars: individual: HD153919 -- 4U1700-37 -- 
Ultraviolet: stars}}

\maketitle

\section{Introduction}

The massive stars in the Milky Way are not randomly distributed, but
are concentrated in loose groups called OB associations located
in the spiral arms of our galaxy (for a review, see e.g.\ Brown et al.
\cite{BB99}). About 80\% of the O stars are member of an OB
association; the kinematical properties of the remaining 20\% of the
field population suggest that these O~stars are runaways, i.e. they
were born in an OB association, but at a certain stage they escaped
from it (Blaauw \cite{Bl93}). The two most popular scenarios to explain
the existence of runaway stars are (i) the dynamical ejection from a
young cluster (Poveda et al. \cite{PR67}) and (ii) the supernova of
the companion star in a massive binary (Blaauw \cite {Bl61}). A recent
study by Hoogerwerf et al. (\cite{HD00}) based on {\it Hipparcos} data
demonstrates that both scenarios are at work, probably at a rate of
1:2, respectively.

High-mass X-ray binaries (HMXBs) are the descendants of massive
binaries (Van den Heuvel \& Heise \cite{HH72}). A neutron star or a
black hole, the compact remnant of the initially most massive star
(the primary) in the binary system, produces X-rays due to the
accretion of matter from the secondary (an OB supergiant or a Be
star); see Kaper (\cite{Ka98}) for an overview of the OB-supergiant
systems. The binary system remains bound after the supernova,
if less than 50\% of the total system mass is lost during the
(assumed symmetric) explosion (Blaauw \cite{Bl61}, Boersma
\cite{Bo61}). The latter can be understood if one considers the phase
of mass transfer occuring when the primary becomes larger than its
critical Roche lobe (e.g. at the end of core-hydrogen burning when the
star expands to become a supergiant) and matter flows from the primary
to the secondary. This results in a change of the mass ratio from
larger to smaller than one. A kick exerted on the compact object due
to the eventual asymmetry of the supernova explosion has also to be
taken into account when determining whether the binary breaks up or
remains bound after the supernova.

\begin{figure}[t]
\centerline{\psfig{figure=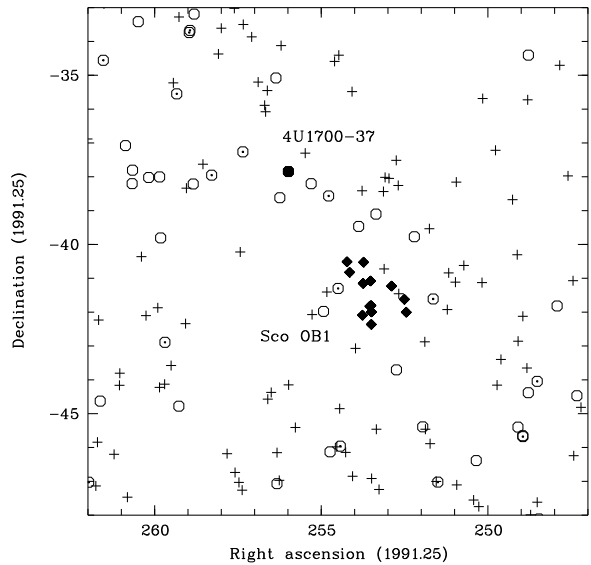,height=8cm,angle=-90}}
\caption[]{O- and B-type stars selected from the {\it Hipparcos}
catalogue in the field of HD153919/4U1700-37 (filled circle). The
confirmed members of Sco~OB1 are shown as filled diamonds. The plus
symbols correspond to OB stars with a parallax larger than 1 mas
(i.e. distance smaller than 1~kpc) and were eliminated from the
membership analysis. The open circles indicate stars with a
(photometric) distance within the range of Sco~OB1; some of them, with
a proper motion similar to the confirmed association members, are
indicated by a circle with central dot (cf.\ Fig.~\ref{fig2}). The
latter close to Sco OB1 might be members as well.
}
\label{fig1}
\end{figure}

\begin{figure}[t]
\centerline{\psfig{figure=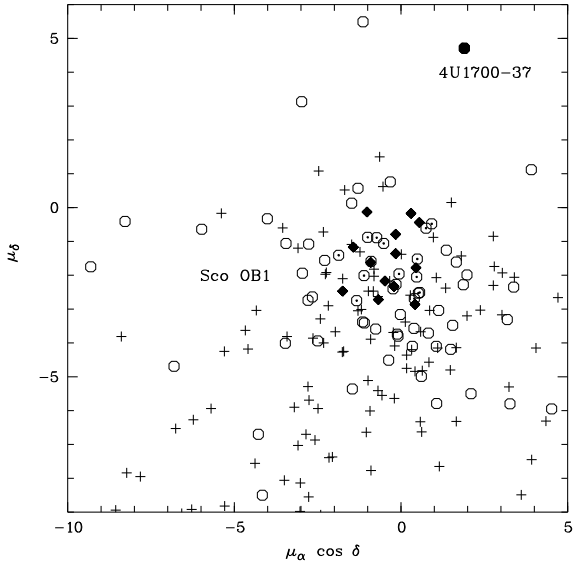,height=8cm,angle=-90}}
\caption[]{The {\it Hipparcos} proper motions of the OB stars shown in
Fig. 1. The filled diamonds represent the confirmed members of
Sco~OB1; these stars cluster both in location and in proper
motion. The plusses and open circles indicate the OB stars we could,
or could not exclude on the basis of a distance estimate,
respectively. The circles with a central dot are stars with similar
proper motion and photometric distance as the confirmed cluster
members. Also shown is 4U1700-37 (filled circle)
which obviously has a
proper motion different from that of Sco~OB1.}
\label{fig2}
\end{figure}

According to the binary-supernova scenario all HMXBs should be
runaways. Gies \& Bolton (\cite{GB86}) did not find observational
evidence supporting this hypothesis on the basis of radial-velocity
measurements, though Van Oijen (\cite{VO89}) found strong indications
that HMXBs are high-velocity objects. Based on pre-{\it Hipparcos}
proper motion measurements, Van Rensbergen et al. (\cite{VV96})
suggested that the HMXB Vela~X-1 is a runaway system produced by the
supernova scenario, and that it originates in the OB association
Vel~OB1. The discovery of a wind-bow shock around Vela~X-1 showed that
this system indeed is running through interstellar space with a
supersonic velocity, proving the runaway nature of this HMXB (Kaper et
al. \cite{KL97}). The {\it Hipparcos} proper motions of a dozen HMXBs
(Chevalier \& Ilovaisky \cite{CI98}, Kaper et al. \cite{KC99}) finally
demonstrated that, as expected, likely all HMXBs are runaways. The
most massive systems (those hosting an OB supergiant) have a mean
peculiar (i.e. with respect to their standard of rest) tangential
velocity of about 40~\kms, whereas the Be/X-ray binaries have on
average lower velocities (about 15~\kms). This difference in velocity
is consistent with the predictions of binary evolution (Van den Heuvel
et al. \cite{VP01}).

The identification of the ``parent'' OB association of a HMXB is
important, because it provides unique constraints on the evolution of
high-mass X-ray binaries. When the system's proper motion and parent
OB association are known, its kinematical age can be derived. The
kinematical age marks the time of the supernova that produced the
compact X-ray source. The distance of a HMXB usually is quite
uncertain (and required to calculate its space velocity), but the
distance to an OB association can be determined with better
accuracy. The space velocity relates to the amount of mass lost from
the system during the supernova explosion (cf.\ Nelemans et
al. \cite{NT99}). The age of the parent OB association should be equal
to the age of the binary system. Consequently, the turn-off mass at
the time of supernova yields the initial mass of the primary. Thus,
this relatively straigthforward observation can be used to determine
the age of the system, the time of supernova of the primary, the
initial mass of the primary, and the amount of mass lost from the
system during the supernova. Combining this information allows one to
put constraints on the initial orbital parameters of the progenitor of
the HMXB and on the evolutionary history of the system.

Here we apply this to the system HD153919/4U1700-37.  HD153919
($m_{V}=6.6$) is the O6.5~Iaf+ companion to 4U1700-37, most likely a
neutron star powered by wind accretion (Jones et al. \cite{JF73},
Haberl et al. \cite{HW89}), although no X-ray pulsations have been
detected (Gottwald et al. \cite{GW86}). According to Brown et
al. (\cite{BW96}) 4U1700-37 is a good candidate for a low-mass black
hole. HD153919 is the hottest OB companion star known in a HMXB;
therefore, the progenitor of 4U1700-37 potentially is a very massive
star. Chevalier \& Ilovaisky (\cite{CI98}) showed that the {\it
Hipparcos} proper motion of HD153919 (5 mas~yr$^{-1}$) corresponds to
a peculiar tangential motion of 57~\kms\ for an adopted distance of
1.7~kpc (Bolton \& Herbst \cite{BH76}), which proves the runaway
nature of the system.

In the following we will use the {\it Hipparcos} data of OB-type stars
in the Sco-Cen region to search for the parent OB association of
4U1700-37. The result will be used to reconstruct the evolutionary
history of the system.

\section{Sco OB1: the parent OB association of 4U1700-37}

\begin{figure*}
\centerline{\psfig{figure=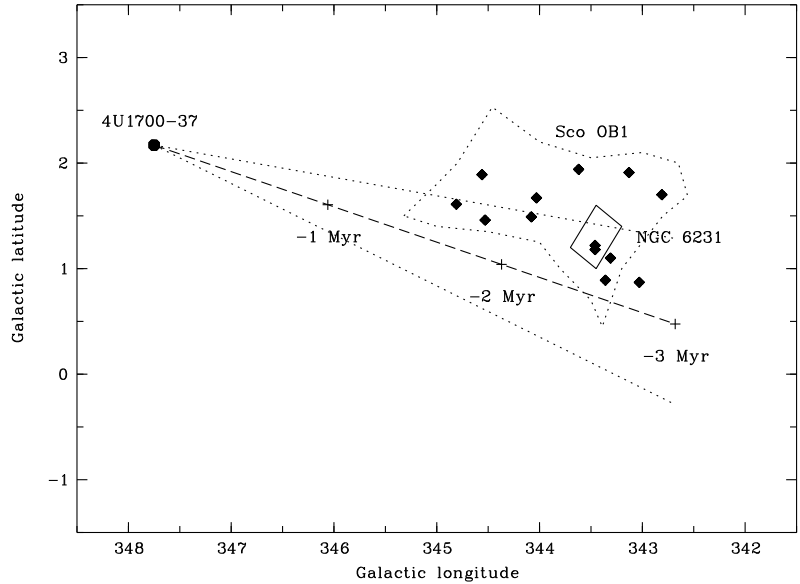,height=12cm,angle=-90}}
\caption[]{The reconstructed path of the runaway HMXB 4U1700-37
intersects with the location of Sco~OB1; the error cone is indicated
by the dotted straight lines. The {\it Hipparcos} confirmed members
are shown as filled diamonds. The dotted line marks the region studied
by Perry et al. (\cite{PH91}), including the young open cluster
NGC~6231 (box). The proper motion of 4U1700-37 is with respect to the
average proper motion of Sco~OB1. The corresponding kinematical age of
4U1700-37 is $2 \pm 0.5$ million year. The current angular separation
between 4U1700-37 and NGC~6231 (at 2~kpc) corresponds to a distance of
about 150~pc.}
\label{fig3}
\end{figure*}

\subsection{Early suggestions}

We now consider in which association 4U1700-37 may have originated.
In their paper on the open cluster NGC~6281, Feinstein \& Forte
(\cite{FF74}) remark that HD153919, a comparison star in their study,
fits remarkably well the color-magnitude and color-color relations of
the open cluster NGC~6231. This cluster, the suggested core of the
Sco~OB1 association, is a few degrees away from NGC~6281. Feinstein \&
Forte suggested that HD153919 may be a runaway star from
NGC~6231. Based on the proper motion listed in the Smithsonian
Astrophysical Observatory catalogue (13 mas~yr$^{-1}$) they obtained a
kinematical age of $1.2 \times 10^{6}$~yr, ``roughly in agreement with
the age of NGC~6231 which is a very young cluster''.

\subsection{Hipparcos observations of the OB stars around 4U1700-37}

In our search for the parent OB association of 4U1700-37 we used the
{\it Hipparcos} database (ESA \cite{ESA97}, Perryman et al. \cite{PL97}) and
selected the OB stars contained in a region of $20 \times 20$ degrees
centered on 4U1700-37 (Fig.~\ref{fig1}). In principle, the {\it
Hipparcos} data (location, magnitude, parallax, and proper motion) of
the OB stars should be sufficient to identify the OB associations in
that area. However, the {\it Hipparcos} data, in particular the
parallax, are only accurate enough for OB stars closer than about 1
kpc (cf.\ De Zeeuw et al. \cite{DH99} for a {\it Hipparcos} census of
the nearby OB associations). The estimated distance of 4U1700-37
(1.7~kpc) indicates that the candidate parent OB association of this
runaway system is not within the required range for an accurate {\it
Hipparcos} cluster membership analysis. In order to identify the
likely members of an OB association we have to rely mainly on the
position and proper motion of the O and B-type stars.

The membership list of Humphreys (\cite{Hu78}), based on
radial-velocity studies, is used as a first indication to locate the
OB associations in the area. It turns out that there is only one good
candidate parent OB association in the backward direction of
4U1700-37: Sco OB1, for which distances are quoted in the range
1.6--2.3~kpc (Perry et al. \cite{PH91}, Sung et al. \cite{SB98}).  To
eliminate some foreground stars we used the {\it Hipparcos}
parallax. We also calculated photometric distances (taking into
account an estimate of the interstellar extinction using the spectral
type) to eliminate stars from the {\it Hipparcos} input list which are
either nearby ($d < 1$ kpc) or far away ($d > 3$ kpc) compared to the
distance of Sco OB1.

We identified several members of Sco~OB1 in the {\it Hipparcos}
catalogue which are also given as members in Humphreys (\cite{Hu78})
and Perry et al. (\cite{PH91}). In Fig.~\ref{fig2} we show the
observed proper motions of the OB stars in the field. Using the mean
location and mean proper motion of the OB stars in common with those
listed in Humphreys and Perry et al., we could identify a few more
candidate OB-type members of Sco~OB1. The ``{\it Hipparcos}
confirmed'' members of Sco~OB1, with spectral type, proper motion, and
radial velocity (from Humpreys \cite{Hu78} and Gies \& Bolton
\cite{GB86}) are listed in Table~\ref{tab1} and are indicated in the
figures with a filled diamond. The mean proper motion of these
members, and thus an estimate of the proper motion of Sco~OB1, is:
$\mu_{\alpha}\cos{\delta}=-0.39$ mas yr$^{-1}$, $\mu_{\delta}=-1.54$
mas yr$^{-1}$. This is in close agreement with the proper motion
expected on the basis of differential galactic rotation and peculiar
solar motion; correction for the latter two effects gives
$\mu_{l}^{\rm pec} = 0.57$~mas~yr$^{-1}$ and $\mu_{b}^{\rm pec} =
-0.10$~mas~yr$^{-1}$, which corresponds to a tangential velocity of
$\sim 5$~\kms\ with respect to its standard of rest. 

\begin{table*}
\caption[]{The confirmed {\it Hipparcos} members of Sco~OB1 and the high-mass
X-ray binary HD153919/4U1700-37.  The columns list the HD number, {\it
Hipparcos} catalogue number HIP, the observed proper motion in right
ascension and declination, the spectral type (Perry et
al. \cite{PH91}), V magnitude, and observed heliocentric radial
velocity (from Humphreys \cite{Hu78}), respectively.}
\begin{flushleft}
\begin{tabular}{llrrllc} \hline
\multicolumn{7}{c}{Members of Sco OB1 ({\it Hipparcos} data)} \\
\hline \hline
HD number & HIP & $\mu_{\alpha}\cos{\delta}$ (error) & $\mu_{\delta}$
(error) & Spectral Type & V & $v_{\rm rad}$ \\ 
        & & (mas yr$^{-1}$) & (mas yr$^{-1}$) & & (mag) & (\kms) \\
\hline
151515 & 82366 & -1.43(0.68) & -1.17(0.56) & O7 II(f) & 7.16 & var. \\
151564 & 82378 & -0.91(0.95) & -1.62(0.68) & B0.5 V & 7.99 & -39.6 \\
152235 & 82669 & -0.21(0.76) & -2.33(0.58) & B0.7 Ia & 6.28 & -36.0 \\
152234 & 82676 & -1.75(1.44) & -2.47(1.02) & B0.5 Ia & 5.46 & -6.0 \\
152246 & 82685 & -0.16(0.92) & -0.79(0.62) & O9 III-IVn & 7.32 & 8.0 var.\\
152405 & 82767 & -1.02(0.88) & -0.13(0.71) & O9.7 Ib-II & 7.20 & -8 \\
152424 & 82783 & -0.68(0.75) & -2.72(0.57) & OC9.7 Ia & 6.30 & -18.0 var.\\
152667 & 82911 & 0.30(0.78) & -0.17(0.65) & B0.5 Ia & 6.18 & -5.0 \\ 
151804 & 82493 & 0.55(0.73) & -0.44(0.54) & O8 Iaf & 5.23 & -61.0 \\
152236 & 82671 & -0.48(0.75) & -2.17(0.61) & B1.5 Ia+p & 4.70 & -23.9 \\
152248 & 82691 & 0.42(1.46) & -2.86(0.96) & O7 Ib:(f) + O6.5:f & 6.07 & -44 \\
152408 & 82775 & -0.16(0.67) & -1.36(0.51) & O8: Iafpe & 5.78 & var. \\
152723 & 82936 & 0.45(1.47) & -1.78(1.04) & O6.5 III(f) & 7.10 & -3.5 \\ \hline
\multicolumn{7}{c}{HD153919/4U1700-37} \\ \hline
153919 & 83499 & 1.90(0.78) & 4.71(0.48) & O6.5Iaf+ & 6.48 & -60.0 \\ \hline
\end{tabular}
\end{flushleft}
\label{tab1}
\end{table*}

\subsection{The kinematical age of 4U1700-37}

Figure~\ref{fig3} displays the {\it Hipparcos} members of Sco~OB1. We
have also indicated the area of Sco~OB1 studied by Perry et
al. (\cite{PH91}; note that the association might well extend beyond
these borders) as well as the location of the open cluster NGC~6231,
the suggested nucleus of Sco~OB1. Subtraction of the average proper
motion of Sco~OB1 from the observed proper motion of HD153919
(4U1700-37) results in the path sketched in Fig.~\ref{fig3}. In
principle, the galactic potential should be taken into account when
reconstructing the path of the runaway system, but for this relatively
short track the corrections will be very small. The uncertainty in the
proper motion measurement of HD153919 (Table~\ref{tab1}) allows for a
range in position represented by the straight dotted lines. Clearly,
4U1700-37 has been within the area of Sco~OB1 about 2 million years
ago; also NGC~6231 is included in the error cone. We derive a
kinematical age of the system of $2 \pm 0.5$ million years. Given the
large proper motion of the system, the kinematical age can be derived
with relatively high precision. Note that this age determination is
independent of the adopted distance to 4U1700-37 and Sco~OB1.
   
\subsection{The distance and age of Sco~OB1}

Perry et al. (\cite{PH91}) determine the distance of Sco~OB1 at
2.0~kpc, very similar to the 1.9~kpc reported by Humphreys
(\cite{Hu78}). At a distance of 2~kpc the relative proper motion of
4U1700-37 with respect to Sco~OB1 corresponds to a tangential velocity
of 58~\kms. Taking into account the radial velocities of the members
of Sco~OB1 (mean velocity $-14$~\kms, though the radial velocities
display a large spread, Humphreys \cite{Hu78}) and of HD153919
($-60$~\kms, Gies \& Bolton \cite{GB86}), this results in a space
velocity of 75~\kms.  As HD153919 is moving towards us, its present
distance is about 100~pc less than Sco~OB1, i.e. 1.9~kpc, in agreement
with its photometric distance. For NGC~6231, the open cluster inside
Sco~OB1 (Fig.~\ref{fig3}), Balona \& Laney (\cite{BL95}) derive a
distance modulus of $11.08 \pm 0.05$~mag, and Sung et
al. (\cite{SB98}) arrive at a very similar result: $11.0 \pm
0.07$~mag, corresponding to a distance of 1.6~kpc. If this is the
appropriate distance of the parent association, the present distance
of HD153919 is about 1.5~kpc, and its space velocity with respect to
NGC~6231 67~\kms. Obviously, NGC~6231 might also be an open cluster in
front of Sco~OB1.

The mean radial velocity of Sco~OB1 of $-14$~\kms\ corresponds to a
distance of 2.0~kpc.  Neutral hydrogen measurements in the direction
of HD153919 by Benaglia \& Cappa (\cite{BC99}) indicate a distance of
2~kpc as well. For the remainder of this paper we adopt a distance of
2~kpc for Sco~OB1.

Based on the evolutionary grids of Maeder \& Meynet (\cite{MM88}),
Perry et al. (\cite{PH91}) derive a logarithmic age of $6.9 \pm 0.2$
(8~Myr) for Sco~OB1 and the open clusters NGC~6231 and Tr~24. For
NGC~6231 Balona \& Laney (\cite{BL95}) estimate an age of $5 \pm
1$~Myr. Using the models of Schaller et al. (\cite{SS92}), Sung et
al. (\cite{SB98}) derive an age of 2.5--4 Myr for the massive stars in
NGC~6231. The low-mass stars in this cluster show a large age
spread. NGC~6231 may represent a relatively young region in
Sco~OB1. Perry et al. (\cite{PH91}) do not find a significant age
difference between Sco~OB1 and the enclosed clusters NGC~6231 and
Tr~24. Anyway, Sco~OB1 certainly is a young OB association given the
large number of O~stars still present.

\section{On the evolutionary history of HD153919/4U1700-37}
 
Our analysis shows that HD153919/4U1700-37 originates in the OB
association Sco~OB1, from which it escaped about 2~Myr ago due to the
supernova of 4U1700-37's progenitor. At the time of the (assumed
symmetric) supernova explosion less than half of the total system mass
was lost from the system, as the system remained bound. The amount of
mass lost during the supernova explosion ($\triangle$M) can be
estimated from the current space velocity $v_{\rm sys}$ of the
system. For a circular pre-supernova orbit and a symmetric supernova
explosion, Nelemans et al. (\cite{NT99}) derive the following relation
between $\triangle$M and $v_{\rm sys}$:
\[
\left( \frac{\triangle {\rm M}}{{\rm M}_{\odot}} \right) = \left(
\frac{v_{\rm sys}}{213 \, {\rm km/s}} \right) \left(
\frac{{\rm M}}{{\rm M}_{\odot}} \right)^{-1} \left( 
\frac{P_{{\rm cir}}}{{\rm day}} \right)^{\frac{1}{3}} \left( \frac{{\rm
M}+m}{{\rm M}_{\odot}} \right)^{\frac{5}{3}} \, ,
\]
where M is the present mass of HD153919, $m$ the mass of 4U1700-37,
and $P_{\rm cir}$ the orbital period after re-circularization of the
orbit due to tidal dissipation. The current orbital period of the
system is 3.41~day and there is no indication that the orbit is
non-circular. As the X-ray source is not pulsating, only the
radial-velocity orbit of the O supergiant can be measured, so that the
masses of both stars are not uniquely determined. Heap \& Corcoran
(\cite{HC92}) propose M~$= 52 \pm 2$~M$_{\odot}$ (i.e. a mass
corresponding to its spectral type) and $m = 1.8 \pm 0.4$~M$_{\odot}$;
Rubin et al. (\cite{RF96}) argue that M~$= 30^{+11}_{-7}$~M$_{\odot}$
and $m = 2.6^{+2.3}_{-1.4}$~M$_{\odot}$. For a space velocity of
75~\kms, $\triangle$M becomes 8~M$_{\odot}$ or 6~M$_{\odot}$ for the
solution of Heap \& Corcoran and Rubin et al.,
respectively. Therefore, the mass of the star that exploded was about
9~M$_{\odot}$\footnote{In case of an asymmetric explosion, the derived
mass would be smaller. However, the kick on the neutron star cannot be
too large, because otherwise the system would have been
disrupted.}. This is significantly higher than model calculations by
e.g.\ Wellstein \& Langer (\cite{WL99}) predict.

What can be said about the initial mass of 4U1700-37's progenitor?
Given its origin in Sco~OB1, the system should have the same age as
the association. As discussed in section 2.4, there likely is some
spread in age within the association, but the observations indicate
that at the moment of the supernova Sco~OB1 was not older than $6
\pm 2$~Myr. The corresponding turn-off mass is 
$\geq 30^{+30}_{-10}$~M$_{\odot}$ (Schaller et al. \cite{SS92}).  Following
Iben \& Tutukov (\cite{IT85}) (and case~B mass transfer), the initial
mass of a star that will explode as a 9~M$_{\odot}$ star is
25~M$_{\odot}$. Although Iben \& Tutukov do not take into account the
mass lost by the helium star, this result is consistent with our
estimate of the progenitor mass based on the age of Sco~OB1.  If we
take the initial mass of the primary to be 30~M$_{\odot}$, the initial
mass of the secondary must have been less, and thus the present mass
of HD153919 cannot be higher than about 60~M$_{\odot}$
(i.e. conservative mass transfer).

It is difficult to reconstruct the evolution of the massive binary
before the supernova explosion. The main problem is the current short
orbital period of the system. Applying Eqs.\ (4) and (5) in Nelemans
et al. (\cite{NT99}), the orbital period before the supernova was a
bit longer, about 4 days. In such a close binary it might well be that
the primary starts transfering mass when it is still on the main
sequence (case~A mass transfer), but then one would predict a
relatively large increase of the orbital period in case of
conservative mass transfer (cf.\ Wellstein \& Langer \cite{WL99}).  It
might be that the evolution has been highly non-conservative due to
strong stellar-wind mass loss and/or non-conservative Roche-lobe
overflow. Case~B mass transfer followed by a contact phase could
produce systems like the Wolf-Rayet binary CQ~Cep/HD214419
(e.g. Marchenko et al. \cite{MM95}) with an orbital period of
1.64~day. The latter system shows that in principle a short-period
system like 4U1700-37's conjectured pre-supernova configuration can be
produced (cf.\ Van den Heuvel \cite{VH73}). A non-conservative
evolutionary scenario for 4U1700-37 is also suggested by Wellstein \&
Langer (\cite{WL99}).

\section{Discussion}

We argue that the initial mass of the progenitor of 4U1700-37 was
$\geq 30^{+30}_{-10}$~M$_{\odot}$. This is relevant for the discussion
which stars leave black holes and which stars end up as neutron
stars. It is commonly believed that the most massive stars form black
holes, while massive stars with a mass below a certain limit (M$_{\rm
BH}$) form neutron stars. This mass limit is under strong debate
(e.g. Ergma \& Van den Heuvel
\cite{EV98}). Maeder (\cite{Ma92}) suggested that the observed helium
and overall metal abundance is best reproduced if M$_{\rm BH} \simeq
20$~M$_{\odot}$, while Timmes et al. (\cite{TW96}) set this limit at
$\sim 30$~M$_{\odot}$.  Whether the mass limit for black-hole
formation in single stars can be compared to that in massive binaries
is not clear (Brown et al. \cite{BW96}). Kaper et al. (\cite{KL95})
set the lower limit for black-hole formation in a massive binary at
$\sim 50$~M$_{\odot}$ based on observations of Wray~977 and X-ray
pulsar companion GX301-2. But Wellstein \& Langer (\cite{WL99})
propose that the initial mass of the neutron star in this system was
much less, about 26~M$_{\odot}$. The same authors derive for single
stars that M$_{\rm BH} \le 25$~M$_{\odot}$.

However, for 4U1700-37 we now have an independent estimate of its
progenitor mass, based on the age of its parent OB association. The
only drawback is that it is not clear whether 4U1700-37 is a neutron
star or a black hole. Up to now, X-ray pulsations, which
would immediately identify the compact star as a neutron star, have
not been detected. The presence of a cyclotron feature in the X-ray
spectrum would also classify the X-ray source as a neutron
star. Reynolds et al. (\cite{RO99}) modeled the X-ray spectrum of
4U1700-37, obtained with {\it BeppoSAX}, and report the presence of a
possible cyclotron feature at an energy of $\sim 37$~keV. If real,
this observation yields a magnetic field strength of about $5 \times
10^{11}$~G, so that 4U1700-37 must be a neutron star. Without
confirmation, the alternative that 4U1700-37 is a low-mass black hole
cannot be excluded. If 4U1700-37 is a neutron star, a lower limit
for black-hole formation in a massive binary derived from this system
would be M$_{\rm BH} = 30^{+30}_{-10}$~M$_{\odot}$.

\begin{acknowledgements}

We thank the referee Dany Vanbeveren for carefully reading the
manuscript.  LK is supported by a fellowship of the Royal Netherlands
Academy of Arts and Sciences. JD acknowledges NWO Spinoza grant 08-0
to E.P.J. van den Heuvel. Ed van den Heuvel and Gijs Nelemans are
thanked for stimulating discussions.

\end{acknowledgements}

\end{document}